\documentclass[11pt,a4paper]{article}%
\usepackage{rotating}
\usepackage{lscape}
\usepackage{amsfonts}
\usepackage{amssymb}
\usepackage[centertags]{amsmath}
\usepackage{graphicx}
\usepackage{adjustbox}
\usepackage{arydshln}
\usepackage{color}%
\providecommand{\U}[1]{\protect\rule{.1in}{.1in}}

\begin{document}

\title{High-Frequency Jump Tests: Which Test Should We Use?\thanks{This research has
been supported by Australian Research Council Discovery Grants No.
{DP150101728 and DP170100729}. We thank a co-editor, an associate editor and
two anonymous referees for very helpful and constructive comments on earlier
drafts of the paper. We are also grateful to John Maheu, Herman van Dijk,
Maria Kalli and Jim Griffin, plus participants at the 11th Annual RCEA
Bayesian Econometric Workshop (Melbourne, 2017), for very helpful comments on
an earlier version of the paper.}}
\author{Worapree Maneesoonthorn\thanks{Melbourne Business School, The University of
Melbourne, Australia.}, Gael M. Martin\thanks{Corresponding author. Department
of Econometrics and Business Statistics, Monash University, Australia. Email:
\textit{gael.martin@monash.edu}.}, Catherine S. Forbes\thanks{Department of
Econometrics and Business Statistics, Monash University, Australia.}}
\maketitle

\begin{abstract}
We conduct an extensive evaluation of price jump tests based on high-frequency
financial data. After providing a concise review of multiple alternative
tests, we document the size and power of all tests in a range of empirically
relevant scenarios. Particular focus is given to the robustness of test
performance to the presence of jumps in volatility and microstructure noise,
and to the impact of sampling frequency. The paper concludes by providing
guidelines for empirical researchers about which test to choose in any given setting.

\bigskip


\bigskip


\end{abstract}


\baselineskip18pt

\section{Introduction}

Extreme movements (or `jumps') in asset prices play an important role in the
tail behaviour of return distributions, with the perceived risk (and, hence,
risk premium) associated with this extreme behaviour differing from that
associated with small and regular movements (see, Bates, 1996, and Duffie
\textit{et al}., 2000, for early illustrations of this point, and Todorov and
Tauchen, 2011, Maneesoonthorn \textit{et al}., 2012, and Bandi and Ren\`{o},
2016, for more recent expositions). Indeed, the modelling of jumps, in both
the price itself and its volatility, has been given particular attention in
the option pricing literature, where the additional risk factors implied by
random jumps have helped explain certain stylized patterns in option-implied
volatility (Merton, 1976; Bates, 2000; Duffie \textit{et al}., Eraker, 2004;
Todorov, 2010; Maneesoonthorn \textit{et al.}; Bandi and Ren\`{o}). Evidence
of price jump clustering in spot returns - whereby price and/or volatility
jumps occur in consecutive time periods - has also been uncovered, with
various approaches having been adopted to model this dynamic behaviour,
including the use of simultaneous price and volatility jumps over time (Chan
and Maheu, 2002; Eraker \textit{et al}., 2003; Maheu and McCurdy, 2004; Fulop
\textit{et al}., 2014; A{\"{\i}}t-Sahalia \textit{et al}., 2015; Bandi and
Ren\`{o}; Maneesoonthorn \textit{et al}., 2017).

Coincident with the trend towards more sophisticated models for asset prices,
the use of high-frequency intraday data to construct nonparametric measures of
asset price variation - including the jump component thereof - has become
wide-spread. Multiple alternative methods are now available to practitioners,
both for testing for jumps and for measuring price variation in the presence
of jumps, with some empirical analyses exploiting such measures in addition
to, or as a replacement of, measurements based on end-of-day prices. (See
Koopman and Scharth, 2013, Christensen \textit{et al.}, 2014, and
Maneesoonthorn \textit{et al.}, 2017, for recent examples, including
references to earlier work.)

This short paper provides the results of an investigation into the relative
accuracy of the many high-frequency price jump tests that are now on offer.
Particular attention is given to the robustness of the tests to the presence
of jumps in volatility, and to the effect of microstructure noise. The impact
of sampling frequency on test performance is also documented. In addition to
enabling key insights to be drawn, the current study also provides a template
for future studies regarding jump tests that may be of interest.

We begin, in Section 2, by providing a very brief review of price jump tests
that have been proposed to date. These methods are grouped into five
categories: 1) those based on the difference between a measure of total
(squared) variation and a jump-robust measure of integrated variation
(Barndorff-Nielsen and Shephard, 2004, 2006; Huang and Tauchen, 2005; Corsi
\textit{et al.}, 2010; Andersen \textit{et al.}, 2012); 2) those that exploit
measures of higher-order variation (A{\"{\i}}t-Sahalia and Jacod, 2009;
Podolskij and Ziggel, 2010); 3) those based on returns, rather than measures
of variation (Andersen \textit{et al.}, 2007; Lee and Mykland, 2008); 4) those
that exploit a variance swap (Jiang and Oomen, 2008); and 5) those that are
expressly designed to mitigate the impact of microstructure noise
(A{\"{\i}}t-Sahalia \textit{et al.}, 2012; Lee and Mykland, 2012). Section 3
makes note of the various tuning components that influence the price jump
tests; with the performance of these tests documented in Section 4. Guidelines
for practitioners are provided in Section 5.

\section{Review of price jump tests\label{alltests}}

Defining $p_{t}=\ln\left(  P_{t}\right)  $ as the natural log of the asset
price, $P_{t}$ at time $t>0$, we begin by assuming the following jump
diffusion process for $p_{t}$,
\begin{equation}
dp_{t}=\mu_{t}dt+\sqrt{V_{t}}dW_{t}^{p}+dJ_{t}^{p}, \label{dpt}%
\end{equation}
where $W_{t}^{p}$ is the Brownian motion, and $dJ_{t}^{p}=Z_{t}^{p}dN_{t}^{p}%
$, with $Z_{t}^{p}$ denoting the random price jump size and $dN_{t}^{p}$ the
increment of a discrete count process, with $P(dN_{t}^{p}=1)=\delta^{p}dt$ and
$P(dN_{t}^{p}=0)=\left(  1-\delta^{p}\right)  dt$.

The aim of a price jump test is to detect the presence of the discontinuous
component, $dJ_{t}^{p}$, and to conclude whether or not $dN_{t}^{p}$ is
non-zero over a particular period. The availability of high-frequency data has
enabled measures of variation - incorporating both the continuous and
discontinuous components of (\ref{dpt}) - to be constructed over a specified
interval of time, e.g. one day, with the statistical properties of such
measures established using in-fill asymptotics. The relevant distributional
results are then utilized in the construction of a price jump test, where the
null hypothesis is usually that the asset price is continuous over the
particular interval under investigation. All tests investigated in this paper
entertain the same null hypothesis of a continuous price path, with the
alternative hypothesis being that the price path contains jumps and, hence, is discontinuous.

This section reviews available tests based on the concepts of, respectively,
squared variation (Section 2.1), higher-order power variation (Section 2.2),
standardized daily returns (Section 2.3), and variance swaps (Section 2.4), as
well as tests that are designed to be robust to microstructure noise (Section
2.5). The detail of eleven specific test statistics (and limiting
distributions) are summarized in Tables \ref{stats1} and \ref{stats2}. The
role of certain tuning parameters, including those appearing in the test
descriptions in Tables \ref{stats1} and \ref{stats2}, is discussed in Section 3.

\subsection{Squared variation\label{2.1}}

The early literature on price jump testing exploits various measures of the
squared variation of the asset price process. In the context of a
continuous-time price process, as defined in (\ref{dpt}), the object of
interest is the difference between total quadratic variation over a discrete
time period (typically a trading day), $Q\mathcal{V}_{t-1,t}=\int_{t-1}%
^{t}V_{s}ds+\sum_{t-1<s\leq t}^{N_{t}^{p}}\left(  Z_{s}^{p}\right)  ^{2},$ and
variation from the continuous component alone, quantified by the integrated
variance, $\mathcal{V}_{t-1,t}=\int_{t-1}^{t}V_{s}ds.$ By definition, the
difference between these two quantities defines the contribution to price
variation of the discontinuous jumps, $\mathcal{J}_{t-1,t}^{2}=\sum_{t-1<s\leq
t}^{N_{t}^{p}}\left(  Z_{s}^{p}\right)  ^{2},$ and price jump test statistics
can thus be constructed from the difference between various \textit{measures}
of $Q\mathcal{V}_{t-1,t}$ and $\mathcal{V}_{t-1,t}$. In Panel A of Table
\ref{stats1}, we provide details of the test proposed by Barndorff-Nielsen and
Shephard (2004, 2006) (also exploited by Huang and Tauchen, 2005), plus an
alternative test proposed by Corsi \textit{et al.} (2010) and two tests
suggested by Andersen \textit{et al.} (2012) (referenced hereafter as BNS, CPR
and MINRV and MEDRV, respectively). In all cases, the test statistic is
constructed using the difference between realized volatility, $RV_{t}%
=\sum\limits_{i=1}^{M}r_{t_{i}}^{2}$, where $r_{t_{i}}=p_{t_{i}}-p_{t_{i-1}}$
denotes the $i^{th}$ of $M$ equally-spaced returns observed during day $t$,
and a chosen measure of integrated variance. All tests are one-sided
upper-tailed tests by construction.

\subsection{Higher-order $\mathcal{P}$-power variation\label{2.2}}

A second class of price jump test exploits the behaviour of higher-order
$\mathcal{P}$-power variation, and estimators thereof. Following
Barndorff-Nielson and Shephard (2004), let an estimator of the $\mathcal{P}%
$-power variation of $p_{t}$ be defined as $\widehat{B}\left(  \mathcal{P}%
,\Delta_{M}\right)  _{t}=%
{\textstyle\sum\limits_{i=1}^{M}}
\left\vert r_{t_{i}}\right\vert ^{\mathcal{P}},$ where $\Delta_{M}=1/M$
denotes the common length of the time intervals between consecutive returns,
and $\mathcal{P}>0.$ The limiting behaviour of this estimator, for different
values of $\mathcal{P}$, sheds light on the different components of the
variation in $p_{t}$. In the case of $\mathcal{P}=2,$ $\widehat{B}\left(
\mathcal{P},\Delta_{M}\right)  _{t}\overset{p}{\longrightarrow}Q\mathcal{V}%
_{t-1,t}$ as $M\rightarrow\infty,$ as is consistent with the distributional
result that $RV_{t}\overset{p}{\longrightarrow}Q\mathcal{V}_{t-1,t}$, as
$M\rightarrow\infty$. For $0<\mathcal{P}<2$,
\begin{equation}
\frac{\Delta_{M}^{1-\mathcal{P}/2}}{m_{\mathcal{P}}}\widehat{B}\left(
\mathcal{P},\Delta_{M}\right)  _{t}\overset{p}{\longrightarrow}A\left(
\mathcal{P}\right)  _{t}\text{ as }M\rightarrow\infty, \label{ap_lim}%
\end{equation}
where $A\left(  \mathcal{P}\right)  _{t}=%
{\displaystyle\int\nolimits_{t-1}^{t}}
\left\vert V_{s}^{1/2}\right\vert ^{\mathcal{P}}ds$ denotes the $\mathcal{P}%
$-power \textit{integrated} variation, $m_{\mathcal{P}}=E\left(  \left\vert
U\right\vert ^{\mathcal{P}}\right)  =\pi^{-1/2}2^{\mathcal{P}/2}\Gamma\left(
\frac{\mathcal{P}+1}{2}\right)  $ and $U$ denotes a standard normal random
variable. In contrast, for $\mathcal{P}>2$, the increments from the jump
component dominate, and the estimator converges in probability to the
$\mathcal{P}$-power \textit{jump} variation, $B\left(  \mathcal{P}\right)
_{t}=%
{\textstyle\sum\limits_{t-1<s\leq t}}
\left\vert dJ_{s}\right\vert ^{\mathcal{P}}.$ If the jump component in
(\ref{dpt}) is not present and $p_{t}$ is continuous as a consequence, then
the limiting result in (\ref{ap_lim}) holds for any $\mathcal{P}>0$.

These limiting results can be used in a variety of ways to detect jumps.
Specifically, A{\"{\i}}t-Sahalia and Jacod (2009) (ASJ, hereafter) compare
$\widehat{B}\left(  \mathcal{P},\Delta_{M}\right)  _{t}$ constructed over two
different sampling intervals, while Podolskij and Ziggel (2010) (PZ,
hereafter) rely on the limiting distribution of a modified version of
$\widehat{B}\left(  \mathcal{P},\Delta_{M}\right)  _{t}$. Both tests are
one-sided, with the ASJ test being lower-tailed, while the PZ test is
upper-tailed. These two approaches are outlined in Panel B of Table
\ref{stats1}.

\subsection{Standardized returns\label{2.3}}

Rather than construct price jump test statistics from various measures of
variation, an alternative is to consider the behaviour of (appropriately
standardized) returns themselves. In brief, based on the assumption of
Brownian motion for the asset price, the return computed over {a chosen}
interval length and scaled by the square root of a consistent estimator of the
corresponding integrated variance, should be asymptotically standard normal if
price jumps are absent. Two tests that exploit this property are proposed by
Andersen \textit{et al.} (2007) and Lee and Mykland (2008), referenced as ABD
and LM, hereafter. ABD conduct multiple two-tailed tests on standardized
returns observed over the trading day, while LM propose an upper-tailed test
based on the maximum absolute standardized return. Details regarding the form
of these two test statistics and their limiting distribution under the null
hypothesis of no jumps are given in Panel A of Table \ref{stats2}.

\subsection{Variance swaps\label{2.4}}

Variance swaps are instruments made up of financial assets and/or derivatives
and are used as tools to hedge against volatility risk. The payoff of a
variance swap can be replicated by taking a short position in the so-called
\textquotedblleft log contract\textquotedblright\ and a long position in the
underlying asset, with the long position being continuously re-balanced (see
Neuberger, 1994). The payoff of such a replicating strategy, computed as the
accumulated difference between proportional returns and continuously
compounded logarithmic returns, equates to half of the integrated variance
when there is no price jump. When a jump is present, the replication error is
completely determined by the realized jump, including the sign of such a jump.
Jiang and Oomen (2008), JO hereafter, exploit this relation in their
two-tailed test construction, with the test details given in Panel B of Table
\ref{stats2}.

\subsection{Treatment for microstructure noise\label{2.5}}

Finally, we investigate the performance of two tests that are specifically
designed to be robust to the presence of microstructure noise. Lee and Mykland
(2012), LM12 hereafter, provide an alternative to the LM test whereby the test
statistic is computed from prices averaged over a small window. A{\"{\i}}%
t-Sahalia \textit{et al.} (2012) propose an alternative test to ASJ (denoted
here by ASJL) that is based on locally smoothed prices. {Similar to their
corresponding predecessors, the LM12 }test, based on extreme value theory,{ is
an upper-tailed }test, while{ the ASJL is a lower-tailed test.} Details of
these two alternative tests are provided in Panel C of Table \ref{stats2}.

\section{Tuning parameter choice\label{tuning}}

As is clear from the outline of the test procedures in Section \ref{alltests},
all require a decision, of one form or another, to be made regarding the
mechanism used to distinguish a continuous increment ($\sqrt{V_{t}}dW_{t}^{p}%
$) from a discontinuous increment ($dJ_{t}^{p}$) in (\ref{dpt}). Such `tuning'
decisions will patently influence the outcome of any test, the values of the
jump measures that are derived from preliminary application of the test and,
hence, any inferential results based on those measures. In this section, we
discuss the impact of alternative choices for the significance level and
certain other tuning values.

\subsection{Significance level\label{3.1}}

All tests are, of course, subject to the selection of a significance level,
which determines the value beyond which the null hypothesis of `no jump' is
rejected. Although certain authors suggest the use of certain (typically
small) significance levels (e.g. Tauchen and Zhou, 2011; ABD), there does not
appear to be widespread consensus in the literature regarding this choice. It
is important to recognize that use of a higher level of significance will
automatically lead to the identification of a greater number of apparent
`jumps', including those having a relatively small magnitude, according to the
usual trade-off between the size of a test and its power in a neighbourhood of
the null hypothesis of no jump. Thus, if the desired focus is to detect (and
subsequently measure) jumps with reasonably large magnitude only, then a small
significance level should be selected. In Section 4 we conduct all size and
power assessments based on two significance levels: 5\% and 1\%.

\subsection{Threshold value\label{3.2}}

The higher-order $\mathcal{P}$-power variation-based PZ, ASJ and ASJL tests,
along with the CPR test, also entail the choice of certain threshold values
that determine the particular jump-free variations that are accumulated. That
is, these threshold values determine whether an individual return belongs to
the diffusive component or to the jump component in the calculation of
$\mathcal{P}$-power variation. The thresholds for these tests are selected as
a multiple ($c_{\vartheta}$) of the local volatility estimate (See Panel A,
Table \ref{stats1}). For example, CPR suggest truncating returns at three
times the local volatility measure. The choice of this multiplier is guided by
the properties of a normal distribution, since if a jump is indeed absent,
then the return (assumed to be normally distributed by approximation of the
diffusive process) should cross the threshold only about 0.3\% of the time.
Naturally, a smaller multiplier will correspond to a larger proportion of
returns being considered as part of the jump contribution. The PZ and ASJ
tests prescribe truncations that involve the choice of the truncation root,
$\varpi$ (see Panel B, Table \ref{stats1}). In both cases, a larger value of
$\varpi$ corresponds to a smaller level of the actual truncation point, again
implying that a larger number of returns are considered as part of the jump
contribution. PZ recommend $\varpi=0.4$, while ASJ recommend the use of
$\varpi=0.48$, recommendations that we adopt for our study of the performance
of these tests, presented in Section 4.

\subsection{Value of $\mathcal{P}$}

The $\mathcal{P}$-power variation-based tests are also subject to the choice
of $\mathcal{P}$ itself, noting that, at least asymptotically, the jump
contribution will dominate for values of $\mathcal{P}>2$. However, with
limited intra-day sampling available, a larger value of $\mathcal{P}$ will
tend to accentuate jumps with relatively large magnitude and thereby diminish
the role in the test outcome of relatively small jumps, for any given choice
of significance level. In Section 4 we report results for the PZ test based on
$\mathcal{P}=2$ and $\mathcal{P}=4$ (PZ2 and PZ4 respectively), to gauge the
effect of this tuning parameter. We also report results for the ASJ and ASJL
tests based on the author-recommended choice of $\mathcal{P}=4$.

\subsection{Sampling interval}

The sampling interval over which intraday returns are computed also plays an
important role in the performance of jump tests. A consensus seems to have
developed in the literature that estimates of $Q\mathcal{V}_{t-1,t}$ and
$\mathcal{V}_{t-1,t}$ are optimally computed over the five-minute interval,
due to the mitigation of microstructure noise when sampling at that
(relatively low) frequency (see Bandi and Russell, 2008). Whether such
optimality carries over to the performance of the corresponding price jump
test is still questionable. It is worth noting that the choice of the sampling
interval will impact each of the ASJ and ASJL tests in two ways - first
through the choice of $\Delta_{M}$, and secondly, via the tuning choice $k$,
which determines the central location of the tests under the null hypothesis.
We conduct both tests using $k=2,$ a choice that is recommended by the
authors. We document test performance over four alternative sampling
frequencies, in order to provide some insight into the influence of the choice
of sampling frequency.

\section{Assessment of test performance \label{Sec:finsample}}

\subsection{Experimental design\label{design}}

We assess the finite sample size and power of each test in empirically
relevant scenarios. In contrast to the earlier assessment of test performance
by Dumitru and Urga (2012), our simulation exercise is used to shed particular
light on the robustness of (an expanded set of) price jump tests to the
presence of a discontinuous volatility process. This is something that has
not, to our knowledge, been documented elsewhere, for any of the tests
discussed, and which is relevant to recent empirical work, in which both
(possibly dynamic) price and volatility jumps are modelled (see Fulop
\textit{et al}., 2014 and Maneesoonthorn \textit{et al}., 2017, for examples).

We generate data from the process in (\ref{dpt}), in conjunction with the
following explicitly defined jump diffusion process for $V_{t}$,
\begin{equation}
dV_{t}=\kappa\left(  V_{t}-\theta\right)  +\sigma_{v}\sqrt{V_{t}}dW_{t}%
^{v}+dJ_{t}^{v}. \label{dvt}%
\end{equation}
The Brownian increment $dW_{t}^{v}$ is assumed to be correlated with
$dW_{t}^{p}$ (in (\ref{dpt})) with $corr\left(  dW_{t}^{v},dW_{t}^{p}\right)
=\rho$, but is assumed to be uncorrelated with the increment in the volatility
jump process, denoted by $dJ_{t}^{v}$. The data are generated using parameter
values: $\mu_{t}=0$, $\kappa=5$, $\theta=0.4^{2}$, $\sigma_{v}=0.5$ and
$\rho=-0.5$ (adhering to the theoretical restriction $2\kappa\theta
\geqslant\sigma_{v}^{2}$), and with the diffusive variance process initialized
at $\theta$.\footnote{This DGP and its parameter settings are also used in the
simulation exercise of A\"{\i}t-Sahalia and Jacod (2009), and broadly reflect
empirical results recorded in the literature. See Eraker \textit{et al.}
(2003) and Fulop \textit{et al.} (2014) for examples.}

A very fine Euler discretization is employed to simulate high-frequency
observations, with 21600 observations created per trading day, equivalent to
generating price observations every one second over a six-hour trading period.
The price jump test statistics are then constructed using four different
sampling frequencies: five seconds, 30 seconds, one minute and five minutes.
We then compute, over 1000 independent Monte Carlo replications, the
proportion of times that a test detects a price jump, under several different
scenarios. First, we set $dJ_{t}^{p}=0$ for all $t$, and assess the size of
the tests, both in the absence ($dJ_{t}^{v}=0$) and in the presence
($dJ_{t}^{v}\neq0$) of volatility jumps. When a volatility jump is present,
only one jump occurs and its arrival time is random. The size of the
volatility jump increment is assumed to be either `moderate', $dJ_{t}%
^{v}=3\theta$, or `large', $dJ_{t}^{v}=10\theta$. Secondly, we assess the
power of the tests in the case where a single price jump arrives randomly over
the day, and where the price jump size is either `moderate', $dJ_{t}%
^{p}=3\sqrt{\theta}$, or `large', $dJ_{t}^{p}=10\sqrt{\theta}$. Power is also
assessed in both the absence and the presence of volatility
jumps.\footnote{The size of the price jump is expressed as a proportion of the
square root of the long run variance ($\theta$), to fit with the scale of
$dp_{t}.$ The size of the volatility jump, on the other hand, is expressed as
a proportion of the long run variance itelf.}

Finally, we assess the size and power of all tests in the presence of
microstructure noise, under both independent and identically distributed
($i.i.d.$) Gaussian and Student-t noise assumptions, as well as when the
microstructure noise follows a Gaussian autocorrelated process
(A\"{\i}t-Sahalia and Mancini, 2008). Test performance in the absence and
presence of microstructure noise is documented in Section \ref{size} and
Section \ref{micronoise} respectively.

\subsection{Test performance in the absence of microstructure
noise\label{size}}

\subsubsection{Empirical size}

In Table \ref{tab:size}, we report empirical size, based on nominal sizes of
5\% and 1\%, with the following summary relevant to the results for both
significance levels.

Two of the four tests based on measures of squared variation, MINRV and MEDRV,
have an empirical size that is closest to the nominal value, for all sampling
frequencies, and in the presence of volatility jumps. In contrast, the CPR
test is uniformly oversized, and the BNS test oversized in the presence of
large volatility jumps, in particular.

The tests based on $\mathcal{P}$-power variation, standardized returns and
variance swaps (PZ2, PZ4, ASJ, ABD, LM and JO) have reasonable size
performance in the absence of volatility jumps, although ASJ and LM are quite
undersized, in particular for the lower sampling frequencies (one and five
minutes). In the presence of volatility jumps, the PZ and ABD tests are very
(at times, grossly) over-sized, whilst the ASJ and LM tests remain somewhat
under-sized\footnote{Our detection of a decreasing test size at lower
frequencies is in line with sizes reported in ASJ's Table 1, notwithstanding
the fact that the lowest frequency they report corresponds to 30 seconds.}. Of
this set, the JO test is the most robust to the presence of volatility jumps,
and to the choice of sampling frequency; but is still not as accurately sized
as MINRV and MEDRV.

In the absence of microstructure noise, the tests designed to accommodate such
noise (LM12 and ASJL) perform well at the highest sampling frequency of five
seconds, and under no volatility jump; but neither test is uniformly robust to
either the sampling frequency or the presence of volatility jumps. It should
be noted that these tests require a smoothing (averaging) process over
sub-blocks throughout the day, so the effective sample size decreases rapidly
as the sampling frequency is reduced.\footnote{In both Lee and Mykland (2012)
and A\"{\i}t-Sahalia \textit{et al.} (2012), test performance is assessed at
extremely high frequency, with the \textit{lowest} frequency recorded being
three seconds for the LM12 test and five seconds for ASJL. Furthermore,
A\"{\i}t-Sahalia \textit{et al.} only assess their test over three consecutive
trading days, as opposed to a single trading day as is typically done in an
empirically relevant context, and as we have done here.}

As a general rule, across all tests, the proportion of incorrect detections of
a price jump increases with size of the volatility jump, highlighting the
confounding influence of this feature of the DGP.

\subsubsection{Empirical power}

Table \ref{tab:power} reports the power of each test conducted at the 5\%
nominal size level.\footnote{Findings at the 1\% significance level were
qualitatively similar and, hence, are not reported.} As is to be expected, all
tests exhibit greater power when the price jump size is larger. However, for
any given price jump size, the level of power still varies, across test,
sampling frequency and volatility jump size. It is interesting to note that
for \textit{all} tests, and for \textit{all} designs, power is greatest when
the test is conducted at the highest frequency.

Except for the LM12 and ASJL tests, power uniformly decreases as the
volatility jump size gets larger, when the test is conducted at the highest
frequency of five seconds. As the sampling interval becomes longer however, a
variety of patterns are observed, in particular when the price jump is only moderate.

Whilst the power of the ASJ test is always highest at the highest frequency
(as tallies with the qualitative finding in Dumitru and Urga, 2012) it ranks
lowest amongst all tests, overall, with a power that does not exceed 50\% over
all designs considered.\footnote{ASJ entertain two different null hypotheses:
one where the price path is continuous under the null, and one where the null
accommodates a discontinuous price path. In the latter case, the ability of
their test to detect a price jump is measured by empirical size (only), and
this is what they report. However, a key assumption underlying the
distribution of the test statistic under this null hypothesis is that there
are no common price and volatility jumps. We view this as a very restrictive
assumption, particularly given the noted interest and empirical evidence found
in support of this situation. Hence our use of the null hypothesis of a
continuous process, and our documentation of empirical power under scenarios
that allow for volatility jumps, including those that occur contemporaneously
with a price jump.} At the other end of the spectrum, the PZ2, PZ4 and ABD
tests tend to have the highest power overall, in particular at the higher
frequencies, and are the most robust to the size of the volatility jump.

The (relatively) well-sized MINRV and MEDRV tests also have high power in the
presence or otherwise of volatility jumps, as long as the price jump is large
and the sampling frequency is high (five seconds). This statement needs to be
qualified somewhat for the lower sampling frequencies. In particular, when the
volatility jump is also large a sampling interval beyond 30 seconds leads to
quite a reduction in power for these two tests, as indeed is a feature for all tests.

The tests designed to cater for microstructure noise (LM12 and ASJL) have high
power in its absence only when the sampling frequency is very high (five
seconds), the price jump to be detected is large, and the (confounding)
volatility jump is not.

\subsection{Test performance in the presence of microstructure
noise\label{micronoise}}

The presence of microstructure noise is known to hamper the quality of
measures constructed from high-frequency data, including any subsequent price
jump tests conducted based on these measures (see Hansen and Lunde, 2006, and
Bandi and Russell, 2008, amongst others). In Table \ref{tab:noise}, the
empirical size and power of the tests conducted at a nominal level of 5\%, is
recorded, under the three assumptions of microstructure noise described in
Section \ref{design}\footnote{Results produced for the 1\% nominal level are
available but are not recorded here.}. Power is assessed for the case where
price jump size is large ($dJ_{t}^{p}=10\sqrt{\theta}$) and volatility jumps
are absent.

As a general observation, the presence of microstructure noise impacts
negatively on the empirical size of the tests that are not expressly designed
to cater for noise, and that impact varies according to the form of noise. For
instance, the Student-t noise leads to the most inaccurate empirical sizes
overall, including the least robustness of size to sampling frequency. Once
again, the size of the MEDRV and MINRV tests tend to be the most robust to
sampling frequency, although under the Student-t and (Gaussian) autocorrelated
noise the tests are extremely undersized when computed using five second data.

Microstructure noise of all forms has arguably less impact on power than it
does on size. Other than ASJ, all tests not designed to be robust to noise
retain high power under Gaussian noise, in particular for the higher
frequencies (five and 30 seconds), remembering that this assessment is now
under large price jumps and zero volatility jumps.

Not surprisingly, the tests that are designed to accommodate microstructure
noise are relatively robust to the type of noise assumed, at the highest
frequency of five seconds, with the appropriate size and high power. However,
the LM12 test is severely oversized, and the AJL test lacks power, when
conducted at a lower frequency, mimicking their performance in the no
microstructure noise case.

\section{Guidelines for Practitioners}

To conclude, we offer practitioners the following \textit{guidelines}:\medskip

\begin{enumerate}
\item The ASJ is the least powerful test overall, over the variety of designs
considered, as well as being consistently undersized. This confirms (albeit in
slightly different scenarios) the findings of Dumitru and Urga (2012). Our
finding holds both in the absence and presence of microstructure noise.

\item If microstructure noise alone is thought to be present, the two tests
designed to cater for that feature - LM12 and ASJL - based on very
high-frequency (five second) data, are the best choice. Importantly, these
tests continue to perform well when microstructure is absent, but only when
the sampling frequency remains very high, the price jump size is large and
volatility jumps are absent.

\item If the data generating process is thought to feature volatility jumps,
we advocate for the use of one of the two squared-variation tests, MINRV or
MEDRV. To balance size and power performance, we also advise computation of
the selected test statistic at a moderate frequency of 30 seconds.
\end{enumerate}

While these guidelines flow from the settings covered in the experiments
summarised in the paper, it is prudent to remind the reader that good (or
poor) test performance under the available settings does not guarantee
similarly good (or poor) performance under alternative simulation designs.
Nevertheless, the approach used here, including the justifications used to determine the guidelines, may serve as a template for investigating the
performance of price jump tests, and consequent price variation measures,
under alternative DGPs and/or volatility jump size settings.

\fontsize{10pt}{1} 

\begin{table}[ptbh]
\caption{{The jump tests outlined in Section 2.1 (squared variation) and
Section 2.2 (higher-order power variation). All test statistics reported have
a $N(0,1)$ limiting distribution under the null hypothesis of no
jump.}\bigskip}%
\label{stats1}
\renewcommand\thetable{1 (Part 1)} \centering
\begin{adjustbox}{width=1.03\textwidth,height=0.44\textheight}
		\begin{tabular}
			[c]{lll}
			& & \\
			\textbf{Test}	& \textbf{Test Statistic} & \\ \hline \hline \\
			\multicolumn{3}{l}{\textbf{Panel A: Tests based on squared variation}}\\ \\
			BNS &
			\begin{tabular}
				[c]{l}%
				$\begin{aligned}
				T_{BNS,t} & = \frac{1-\frac{BV_{t}}{RV_{t}}}{\sqrt{\left(  \left(  \frac{\pi}%
						{2}\right)  ^{2}+\pi-5\right)  M^{-1}\max\left(  1,\frac{TP_{t}}{BV_{t}^{2}%
						}\right)  }}, \mbox{ where } BV_{t} = \frac{\pi}{2}\left(  \frac{M}{M-1}\right)
				{\textstyle\sum\limits_{i=2}^{M}}
				\left\vert r_{t_{i}}\right\vert \left\vert r_{t_{i-1}}\right\vert, \\
				TP_{t} & = \mu_{4/3}^{-3}\left(  \frac{M^{2}}{M-2}\right)  \sum_{i=3}^{M}\mid
				r_{t_{i-2}}\mid^{4/3}\mid r_{t_{i-1}}\mid^{4/3}\mid r_{t_{i}}\mid^{4/3}, \mbox{ and }
				\mu_{4/3} = 2^{2/3}\Gamma(7/6)\Gamma(1/2)^{-1}
				\end{aligned}$
			\end{tabular}
			& \\ \\ \hdashline		\\	
			CPR &
			\begin{tabular}
				[c]{l}%
				$\begin{aligned}
				T_{CPR,t} = & \frac{1-\frac{CTBV_{t}}{RV_{t}}}{\sqrt{\left(  \left(  \frac{\pi
						}{2}\right)  ^{2}+\pi-5\right)  M^{-1}\max\left(  1,\frac{CTriPV_{t}}%
						{CTBV_{t}^{2}}\right)  }}, \mbox{ where } CTBV_{t} =\frac{\pi}{2}\left(  \frac{M}{M-1}\right)  \sum\limits_{i=2}^{M}\tau_{1,t_{i}}\tau_{1,t_{i-1}}, \\
				CTriPV_{t} & = \mu_{4/3}^{-3}\left(  \frac{M^{2}}{M-2}\right)  \sum
				\limits_{i=3}^{M}\tau_{4/3,t_{i}}\tau_{4/3,t_{i-1}}\tau_{4/3,t_{i-2}}, \hspace{.3cm}
				\tau_{1,t_{i}} = \left\{
				\begin{array}
				[c]{c}%
				\left\vert r_{t_{i}}\right\vert \text{ for\ }r_{t_{i}}^{2}<\vartheta_{t_{i}}\\ \hspace{1.2cm}
				1.094\sqrt{\vartheta_{t_{i}}}\text{\ for }r_{t_{i}}^{2}>\vartheta_{t_{i}}%
				\end{array}
				\right. , \\
				\tau_{4/3,t_{i}} & = \left\{
				\begin{array}
				[c]{c}%
				\left\vert r_{t_{i}}\right\vert ^{4/3}\text{ for }r_{t_{i}}^{2}<\vartheta
				_{t_{i}}\\
				1.129\vartheta_{t_{i}}^{2/3}\text{ for\ }r_{t_{i}}^{2}>\vartheta_{t_{i}}%
				\end{array}
				\right. , \hspace{.2cm}\mbox{ and } \vartheta_{t_{i}} = c_{\vartheta}^{2}\widehat{V}_{t_{i}},
				\mbox{where } \widehat{V}_{t_{i}} \mbox{ denotes a local variance estimator}
				\end{aligned}$
			\end{tabular}
			& \\ \\ \hdashline \\
		MINRV &
			\begin{tabular}
				[c]{l}%
				$T_{MinRV,t}=\frac{1-\frac{MinRV_{t}}{RV_{t}}}{\sqrt{1.81M^{-1}\max\left(
						1,\frac{MinRQ_{t}}{MinRV_{t}^{2}}\right)  }}$, where $MinRV_{t}=\frac{\pi}%
				{\pi-2}\left(  \frac{M}{M-1}\right)  \sum_{i=2}^{M}\min\left(  \left\vert
				r_{t_{i}}\right\vert ,\left\vert r_{t_{i-1}}\right\vert \right)  ^{2}$\\
				and $MinRQ_{t}=\frac{\pi}{3\pi-8}\left(  \frac{M^{2}}{M-1}\right)  \sum
				_{i=2}^{M}\min\left(  \left\vert r_{t_{i}}\right\vert ,\left\vert r_{t_{i-1}%
				}\right\vert \right)  ^{4}$%
			\end{tabular}
			& \\ \\ \hdashline \\
		MEDRV &
			\begin{tabular}
				[c]{l}%
				$T_{MedRV,t}=\frac{1-\frac{MedRV_{t}}{RV_{t}}}{\sqrt{0.96M^{-1}\max\left(
						1,\frac{MedRQ_{t}}{MedRV_{t}^{2}}\right)  }}$, where $MedRV_{t}=\frac{\pi}%
				{\pi+6-4\sqrt{3}}\left(  \frac{M}{M-2}\right)  \sum_{i=3}^{M}med\left(
				\left\vert r_{t_{i}}\right\vert ,\left\vert r_{t_{i-1}}\right\vert ,\left\vert
				r_{t_{i-2}}\right\vert \right)  ^{2}$\\
				and $MedRQ_{t}=\frac{3\pi}{9\pi+72-52\sqrt{3}}\left(  \frac{M^{2}}%
				{M-2}\right)  \sum_{i=3}^{M}med\left(  \left\vert r_{t_{i}}\right\vert
				,\left\vert r_{t_{i-1}}\right\vert ,\left\vert r_{t_{i-2}}\right\vert \right)
				^{4}$%
			\end{tabular}
			& \\ \\ \hline \\
			\multicolumn{3}{l}{\textbf{Panel B: Tests based on $\mathcal{P}$-power variation}}\\ \\
			PZ &
			\begin{tabular}
				[c]{l}%
				$T_{PZ,t}=\frac{M^{\frac{\mathcal{P}-1}{2}}%
					{\textstyle\sum\limits_{i=1}^{M}}
					\left\vert r_{t_{i}}\right\vert ^{\mathcal{P}}\left(  1-\eta_{i}%
					\mathbf{1}_{\left\{  \left\vert r_{t_{i}}\right\vert <\vartheta\left(
						\Delta_{M}\right)  ^{\varpi}\right\}  }\right)  }{\sqrt{Var\left(  \eta
						_{i}\right)  M^{\frac{\mathcal{P}}{2}-1}%
						{\textstyle\sum\limits_{i=1}^{M}}
						\left\vert r_{t_{i}}\right\vert ^{\mathcal{P}}\mathbf{1}_{\left\{  \left\vert
							r_{t_{i}}\right\vert <\vartheta\left(  \Delta_{M}\right)  ^{\varpi}\right\}
				}}}$ where $\eta_{i}$ is a symmetric IID random variable\\
				with $E\left(  \eta_{i}\right)  =1$, $Var\left(  \eta_{i}\right)  <\infty$ and
				$E\left(  \left\vert \eta_{i}\right\vert ^{2+d}\right)  >0$ for some $d>0$;
				and $\mathcal{P\geq}2$.
			\end{tabular}
			& \\ \\ \hdashline \\
			ASJ &
			\begin{tabular}
				[c]{l}%
				Under $H_0$ of continuous price path \\
				$T_{ASJ,t}=\left(  \widehat{\Sigma}_{M,t}^{c}\right)  ^{-1/2}\left(
				\widehat{S}\left(  \mathcal{P},k,\Delta_{M}\right)  _{t}-k^{\frac{\mathcal{P}%
					}{2}-1}\right)  $, where $\widehat{S}\left(  \mathcal{P},k,\Delta_{M}\right)
				_{t}=\frac{\widehat{B}\left(  \mathcal{P},k\Delta_{M}\right)  _{t}%
				}{\widehat{B}\left(  \mathcal{P},\Delta_{M}\right)  _{t}}$,  $\widehat{\Sigma}_{M,t}^{c}=\frac{\Delta_{M}M\left(  \mathcal{P},k\right)
					\widehat{A}\left(  2\mathcal{P},\Delta_{M}\right)  _{t}}{\widehat{A}\left(
					\mathcal{P},\Delta_{M}\right)  _{t}^{2}}$\\
				and $\widehat{A}\left(
				\mathcal{P},\Delta_{M}\right)  _{t}=\frac{\Delta_{M}^{1-\mathcal{P}/2}%
				}{m_{\mathcal{P}}}%
				{\textstyle\sum\limits_{i=1}^{M}}
				\left\vert r_{t_{i}}\right\vert ^{\mathcal{P}}\mathbf{1}_{\left\{  \left\vert
					r_{t_{i}}\right\vert <\vartheta\Delta_{M}^{\varpi}\right\}  }$,
				for $k\geq2$ and $\mathcal{P}>2$. \\
				Here, $M(\mathcal{P},k)$ and $m_\mathcal{P}$ are constants defined by expectations of absolute power of standard normal variables.
				
			\end{tabular}
			& \\ \\ 
			\hline \hline
		\end{tabular}
	\end{adjustbox}
\end{table}


\begin{table}[ptbh]
\caption{{The jump tests outlined in Section 2.3 (standardized daily returns),
Section 2.4 (variance swaps) and Section 2.5 (microstructure noise). The
limiting distribution for each test statistic under the null hypothesis of no
price jump given in column 3 below.}\bigskip}%
\label{stats2}
\renewcommand\thetable{1 (Part 2)} \centering
\begin{adjustbox}{width=1.03\textwidth,height=0.33\textheight}
		\begin{tabular}
			[c]{lll}
			& & \\
			\textbf{Test} & \textbf{Test Statistics} & \textbf{Limiting Dist.}\\ \hline \hline \\
			\multicolumn{3}{l}{\textbf{Panel A: Tests based on standardized returns}}\\ \\
			ABD & $T_{ABD,t_{i}}=\frac{r_{t_{i}}}{\sqrt{M^{-1}BV_{t}}}$. The significance
			level needs to be adjusted for multiple testing. & N(0,1)\\ \\ \hdashline \\
			LM &
			\begin{tabular}
				[c]{l}%
				$T_{LM,t}\mathbf{=}\frac{\left(  \max\left(  \widetilde{T}_{LM,t_{i}}\right)
					-C_{M}\right)  }{S_{M}}$ where $\widetilde{T}_{LM,t_{i}}=\frac{\left\vert
					r_{t_{i}}\right\vert }{\sqrt{\hat{V}_{t_{i}}}},$\\ $C_{M}=\frac{\left(  2\log
					M\right)  ^{1/2}}{0.8}-\frac{\log\pi+\log(\log M)}{1.6\left(  2\log\pi\right)
					^{1/2}}$,\\
				$S_{M}=\frac{1}{0.6\left(  2\log\pi\right)  ^{1/2}}$ and $\hat{V}_{t_{i}}$
				denotes the local variance estimate
			\end{tabular}
			& Gumbel\\ \\ \hline \\
			\multicolumn{3}{l}{\textbf{Panel B: Test based on variance swap}}\\
			JO &
			\begin{tabular}
				[c]{l}%
				$T_{JO,t}=\frac{BV_{t}}{M^{-1}\sqrt{\widehat{\Omega}_{SwV}}}\left(
				1-\frac{RV_{t}}{SwV_{t}}\right)  $, where $SwV_{t}=2\sum_{i=1}^{M}\left(
				R_{t_{i}}-r_{t_{i}}\right)  $ \\ with $R_{t_{i}}$ = arithmatic returns\\
				and $\widehat{\Omega}_{SwV}=3.05\frac{M^{3}}{M-3}\sum\limits_{i=1}^{M}%
				\prod\limits_{k=0}^{3}\left\vert r_{t_{i-k}}\right\vert ^{3/2}$%
			\end{tabular}
			& N(0,1)\\ \\ \hline \\
			\multicolumn{3}{l}{\textbf{Panel C: Tests that account for microstructure noise}}\\
			LM12 &
			\begin{tabular}
				[c]{l}%
				$T_{LM12}=\underset{t_{j}\in G_{n}^{k}}{\max}\frac{\left\vert \chi\left(
					t_{j}\right)  \right\vert -A_{n}}{B_{n}}$ where $A_{n}=\left(  2\log
				\lfloor\frac{n}{kM}\rfloor\right)  ^{1/2}-\frac{\log\pi+\log(\log\lfloor
					\frac{n}{kM}\rfloor)}{2\left(  2\log\lfloor\frac{n}{kM}\rfloor\right)  ^{1/2}%
				},$\\
				$B_{n}=\left(  2\log\lfloor\frac{n}{kM}\rfloor\right)  ^{1/2},$ $\chi\left(
				t_{j}\right)  =\sqrt{\frac{M}{V_{n}}}\left(  \widehat{p}_{t_{j}+kM}%
				-\widehat{p}_{t_{j}}\right)  $, \\ and $\widehat{p}_{t_{j}}$ is the average log
				price over a block size $M$, \\ with the average price computed using every $k^{th}$
				observations. \\
				$V_{n}$ is computed using Podolskij \&\ Vetter (2009)
			\end{tabular}			& Gumbel\\ \\ \hdashline \\
			ASJL &
			\begin{tabular}
				[c]{l}%
				Test statistic as in ASJ, but with the power variation computed \\from smoothed
				log prices. The estimator of the asymptotic variance \\of the test statistic is modified
				accordingly.\\
				Code to conduct the test is available at https://sites.duke.edu/jiali/research/.
			\end{tabular}
			& N(0,1) \\ \\ \hline \hline
		\end{tabular}
	\end{adjustbox}
\end{table}


\begin{table}[ptbh]
\caption{The empirical size of price jump tests constructed at the five
second, 30 second, one minute and five minute sampling frequencies. The tests
are conducted under three DGPs. DGP1 assumes volatility jumps are absent
(labelled as `$dJ_{t}^{v}=0$'); DGP2 assumes that a volatility jump is present
and is of moderate size (labelled as `$dJ_{t}^{v}=3\theta$'); DGP3 assumes a
large volatility jump is present (labelled as `$dJ_{t}^{v}=10\theta$').
Nominal sizes of 5\% and 1\% are shown in columns 3-6 and 7-10, respectively.
Microstructure noise is absent. \bigskip}%
\label{tab:size}
\renewcommand\thetable{2} \centering
\begin{tabular}
[c]{rlrrrr|rrrr}
&  & \multicolumn{4}{c|}{Nominal Size = 5\%} & \multicolumn{4}{c}{Nominal Size
= 1\%}\\
&  & \multicolumn{1}{l}{5 sec} & \multicolumn{1}{l}{30sec} &
\multicolumn{1}{l}{1min} & \multicolumn{1}{l|}{5min} & \multicolumn{1}{l}{5
sec} & \multicolumn{1}{l}{30sec} & \multicolumn{1}{l}{1min} &
\multicolumn{1}{l}{5min}\\\hline\hline
&  &  &  &  &  &  &  &  & \\
& BNS & 0.051 & 0.052 & 0.042 & 0.055 & 0.008 & 0.009 & 0.010 & 0.020\\
& CPR & 0.086 & 0.133 & 0.165 & 0.464 & 0.016 & 0.045 & 0.059 & 0.241\\
& MINRV & 0.051 & 0.046 & 0.048 & 0.044 & 0.006 & 0.005 & 0.005 & 0.010\\
& MEDRV & 0.054 & 0.041 & 0.057 & 0.060 & 0.010 & 0.008 & 0.005 & 0.014\\
\multicolumn{1}{l}{DGP1} & PZ2 & 0.047 & 0.057 & 0.070 & 0.092 & 0.009 &
0.018 & 0.028 & 0.051\\
\multicolumn{1}{l}{no VJ} & PZ4 & 0.046 & 0.051 & 0.058 & 0.078 & 0.014 &
0.015 & 0.026 & 0.046\\
\multicolumn{1}{l}{($dJ^{v}_{t}=0$)} & ASJ & 0.018 & 0.011 & 0.004 & 0.000 &
0.002 & 0.000 & 0.000 & 0.000\\
& ABD & 0.086 & 0.058 & 0.053 & 0.063 & 0.020 & 0.012 & 0.011 & 0.020\\
& LM & 0.032 & 0.009 & 0.007 & 0.005 & 0.003 & 0.001 & 0.002 & 0.000\\
& JO & 0.054 & 0.049 & 0.068 & 0.079 & 0.007 & 0.011 & 0.016 & 0.025\\
& LM12 & 0.037 & 0.124 & 0.209 & 0.521 & 0.011 & 0.055 & 0.103 & 0.386\\
& ASJL & 0.055 & 0.057 & 0.011 & 0.093 & 0.018 & 0.009 & 0.002 & 0.091\\\hline
&  &  &  &  &  &  &  &  & \\
& BNS & 0.078 & 0.078 & 0.066 & 0.066 & 0.021 & 0.030 & 0.018 & 0.022\\
& CPR & 0.104 & 0.127 & 0.155 & 0.275 & 0.033 & 0.045 & 0.048 & 0.105\\
& MINRV & 0.050 & 0.057 & 0.049 & 0.049 & 0.014 & 0.021 & 0.007 & 0.010\\
& MEDRV & 0.054 & 0.056 & 0.044 & 0.061 & 0.011 & 0.017 & 0.010 & 0.021\\
\multicolumn{1}{l}{DGP2} & PZ2 & 0.216 & 0.268 & 0.271 & 0.179 & 0.185 &
0.230 & 0.237 & 0.152\\
\multicolumn{1}{l}{moderate VJ} & PZ4 & 0.214 & 0.260 & 0.268 & 0.180 &
0.189 & 0.229 & 0.233 & 0.150\\
\multicolumn{1}{l}{($dJ^{v}_{t}=3\theta$)} & ASJ & 0.022 & 0.004 & 0.004 &
0.000 & 0.001 & 0.000 & 0.000 & 0.000\\
& ABD & 0.827 & 0.532 & 0.409 & 0.194 & 0.568 & 0.259 & 0.212 & 0.083\\
& LM & 0.023 & 0.009 & 0.009 & 0.011 & 0.003 & 0.000 & 0.000 & 0.001\\
& JO & 0.053 & 0.068 & 0.078 & 0.110 & 0.011 & 0.017 & 0.021 & 0.045\\
& LM12 & 0.205 & 0.287 & 0.372 & 0.579 & 0.079 & 0.156 & 0.224 & 0.438\\
& ASJL & 0.113 & 0.088 & 0.035 & 0.118 & 0.041 & 0.030 & 0.007 & 0.111\\\hline
&  &  &  &  &  &  &  &  & \\
& BNS & 0.121 & 0.119 & 0.115 & 0.101 & 0.040 & 0.060 & 0.045 & 0.048\\
& CPR & 0.140 & 0.153 & 0.157 & 0.192 & 0.053 & 0.079 & 0.065 & 0.093\\
& MINRV & 0.048 & 0.072 & 0.054 & 0.068 & 0.016 & 0.023 & 0.012 & 0.023\\
& MEDRV & 0.056 & 0.063 & 0.050 & 0.084 & 0.012 & 0.018 & 0.017 & 0.029\\
\multicolumn{1}{l}{DGP3} & PZ2 & 0.800 & 0.723 & 0.636 & 0.421 & 0.794 &
0.713 & 0.624 & 0.400\\
\multicolumn{1}{l}{large VJ} & PZ4 & 0.802 & 0.718 & 0.635 & 0.424 & 0.794 &
0.711 & 0.619 & 0.399\\
\multicolumn{1}{l}{($dJ^{v}_{t}=10\theta$)} & ASJ & 0.018 & 0.002 & 0.004 &
0.000 & 0.000 & 0.000 & 0.000 & 0.000\\
& ABD & 0.996 & 0.922 & 0.818 & 0.477 & 0.975 & 0.765 & 0.585 & 0.272\\
& LM & 0.031 & 0.033 & 0.030 & 0.042 & 0.005 & 0.008 & 0.006 & 0.016\\
& JO & 0.059 & 0.070 & 0.087 & 0.140 & 0.016 & 0.018 & 0.031 & 0.066\\
& LM12 & 0.544 & 0.535 & 0.595 & 0.703 & 0.324 & 0.379 & 0.433 & 0.581\\
& ASJL & 0.220 & 0.172 & 0.083 & 0.165 & 0.130 & 0.077 & 0.024 &
0.155\\\hline\hline
\end{tabular}
\end{table}

\begin{table}[ptbh]
\caption{The empirical power of price jump tests constructed at the five
second, 30 second, one minute and five minute sampling frequencies. The tests
are conducted under two scenarios for the price jump size: moderate (left
panel, with $dJ^{p}_{t}=3\sqrt{\theta}$) and large (right panel, with
$dJ^{p}_{t} =10\sqrt{\theta}$), and under three DGPs for the volatility jump
process. DGP1 assumes volatility jumps are absent (labelled as `$dJ_{t}^{v}%
=0$'); DGP2 assumes that a volatility jump is present and is of moderate size
(labelled as `$dJ_{t}^{v}=3\theta$'); DGP3 assumes a large volatility jump is
present (labelled as `$dJ_{t}^{v}=10\theta$'). All tests are conducted using a
nominal size of 5\%. Microstructure noise is absent. \bigskip}%
\label{tab:power}
\renewcommand\thetable{3} \centering
\begin{tabular}
[c]{rlrrrr|rrrr}
&  & \multicolumn{4}{c|}{Moderate Price Jump} & \multicolumn{4}{c}{Large Price
Jump}\\
&  & \multicolumn{4}{c|}{($dJ^{p}_{t}=3\sqrt{\theta}$)} &
\multicolumn{4}{c}{($dJ^{p}_{t}=10\sqrt{\theta}$)}\\
&  & \multicolumn{1}{l}{5 sec} & \multicolumn{1}{l}{30sec} &
\multicolumn{1}{l}{1min} & \multicolumn{1}{l|}{5min} & \multicolumn{1}{l}{5
sec} & \multicolumn{1}{l}{30sec} & \multicolumn{1}{l}{1min} &
\multicolumn{1}{l}{5min}\\\hline\hline
&  &  &  &  &  &  &  &  & \\
& BNS & 0.793 & 0.201 & 0.120 & 0.068 & 1.000 & 1.000 & 0.998 & 0.573\\
& CPR & 0.881 & 0.384 & 0.311 & 0.479 & 1.000 & 1.000 & 1.000 & 0.862\\
& MINRV & 0.540 & 0.149 & 0.101 & 0.050 & 1.000 & 1.000 & 0.992 & 0.490\\
& MEDRV & 0.776 & 0.219 & 0.124 & 0.064 & 1.000 & 1.000 & 1.000 & 0.705\\
\multicolumn{1}{l}{DGP1} & PZ2 & 1.000 & 0.813 & 0.375 & 0.098 & 1.000 &
1.000 & 1.000 & 0.961\\
\multicolumn{1}{l}{no VJ} & PZ4 & 1.000 & 0.811 & 0.370 & 0.091 & 1.000 &
1.000 & 1.000 & 0.960\\
\multicolumn{1}{l}{($dJ^{v}_{t}=0$)} & ASJ & 0.492 & 0.191 & 0.050 & 0.000 &
0.492 & 0.487 & 0.484 & 0.017\\
& ABD & 1.000 & 0.907 & 0.486 & 0.079 & 1.000 & 1.000 & 1.000 & 0.968\\
& LM & 0.646 & 0.073 & 0.019 & 0.002 & 1.000 & 0.936 & 0.730 & 0.108\\
& JO & 0.999 & 0.334 & 0.178 & 0.104 & 1.000 & 1.000 & 1.000 & 0.818\\
& LM12 & 0.626 & 0.244 & 0.245 & 0.508 & 1.000 & 0.960 & 0.931 & 0.675\\
& AJL & 0.130 & 0.059 & 0.014 & 0.083 & 0.998 & 0.813 & 0.511 & 0.112\\\hline
&  &  &  &  &  &  &  &  & \\
& BNS & 0.474 & 0.158 & 0.099 & 0.087 & 1.000 & 0.995 & 0.928 & 0.363\\
& CPR & 0.579 & 0.254 & 0.190 & 0.294 & 1.000 & 0.999 & 0.965 & 0.605\\
& MINRV & 0.284 & 0.111 & 0.066 & 0.061 & 1.000 & 0.976 & 0.826 & 0.277\\
& MEDRV & 0.418 & 0.130 & 0.077 & 0.064 & 1.000 & 0.997 & 0.941 & 0.415\\
\multicolumn{1}{l}{DGP2} & PZ2 & 1.000 & 0.598 & 0.338 & 0.204 & 1.000 &
1.000 & 1.000 & 0.806\\
\multicolumn{1}{l}{moderate VJ} & PZ4 & 1.000 & 0.591 & 0.335 & 0.189 &
1.000 & 1.000 & 1.000 & 0.800\\
\multicolumn{1}{l}{($dJ^{v}_{t}=3\theta$)} & ASJ & 0.475 & 0.027 & 0.006 &
0.000 & 0.492 & 0.471 & 0.389 & 0.003\\
& ABD & 1.000 & 0.808 & 0.515 & 0.214 & 1.000 & 1.000 & 1.000 & 0.841\\
& LM & 0.457 & 0.052 & 0.013 & 0.006 & 0.995 & 0.785 & 0.547 & 0.073\\
& JO & 0.818 & 0.138 & 0.118 & 0.121 & 1.000 & 1.000 & 0.989 & 0.524\\
& LM12 & 0.316 & 0.302 & 0.369 & 0.566 & 0.998 & 0.859 & 0.801 & 0.639\\
& AJL & 0.152 & 0.092 & 0.033 & 0.120 & 0.980 & 0.539 & 0.259 & 0.122\\\hline
&  &  &  &  &  &  &  &  & \\
& BNS & 0.283 & 0.160 & 0.117 & 0.119 & 1.000 & 0.881 & 0.666 & 0.237\\
& CPR & 0.322 & 0.211 & 0.165 & 0.214 & 1.000 & 0.905 & 0.745 & 0.368\\
& MINRV & 0.137 & 0.088 & 0.051 & 0.080 & 0.996 & 0.589 & 0.371 & 0.162\\
& MEDRV & 0.173 & 0.098 & 0.064 & 0.081 & 1.000 & 0.787 & 0.530 & 0.198\\
\multicolumn{1}{l}{DGP3} & PZ2 & 0.995 & 0.739 & 0.653 & 0.416 & 1.000 &
1.000 & 0.997 & 0.609\\
\multicolumn{1}{l}{large VJ} & PZ4 & 0.995 & 0.742 & 0.652 & 0.415 & 1.000 &
1.000 & 0.997 & 0.607\\
\multicolumn{1}{l}{($dJ^{v}_{t}=10\theta$)} & ASJ & 0.118 & 0.002 & 0.000 &
0.000 & 0.492 & 0.318 & 0.101 & 0.000\\
& ABD & 1.000 & 0.940 & 0.823 & 0.465 & 1.000 & 1.000 & 0.999 & 0.664\\
& LM & 0.376 & 0.085 & 0.041 & 0.042 & 0.940 & 0.625 & 0.449 & 0.094\\
& JO & 0.180 & 0.087 & 0.096 & 0.157 & 1.000 & 0.923 & 0.640 & 0.236\\
& LM12 & 0.541 & 0.496 & 0.588 & 0.696 & 0.971 & 0.742 & 0.718 & 0.682\\
& ASJL & 0.240 & 0.160 & 0.079 & 0.164 & 0.779 & 0.278 & 0.128 &
0.160\\\hline\hline
\end{tabular}
\end{table}

\begin{table}[ptbh]
\caption{The empirical size and power of price jump tests constructed at the
five second, 30 second, one minute and five minute sampling frequencies, in
the presence of microstructure noise. The tests are conducted under DGP1, but
with three different types of microstructure noise: Gaussian noise; Student-t
noise and autocorrelated noise. Each test is conducted using a nominal size of
5\%, with the power assessment conducted using a large price jump size $dJ^{p}
_{t}=10\sqrt{\theta}$. Volatility jumps are absent. \bigskip}%
\label{tab:noise}
\renewcommand\thetable{4} \centering
\begin{tabular}
[c]{rlrrrr|rrrr}
&  & \multicolumn{4}{c|}{Size} & \multicolumn{4}{c}{Power}\\
&  & \multicolumn{1}{l}{5 sec} & \multicolumn{1}{l}{30sec} &
\multicolumn{1}{l}{1min} & \multicolumn{1}{l|}{5min} & \multicolumn{1}{l}{5
sec} & \multicolumn{1}{l}{30sec} & \multicolumn{1}{l}{1min} &
\multicolumn{1}{l}{5min}\\\hline\hline
&  &  &  &  &  &  &  &  & \\
& BNS & 0.000 & 0.040 & 0.048 & 0.060 & 0.999 & 1.000 & 0.997 & 0.533\\
& CPR & 0.000 & 0.109 & 0.181 & 0.443 & 0.999 & 1.000 & 0.999 & 0.847\\
& MINRV & 0.054 & 0.056 & 0.064 & 0.068 & 0.869 & 0.999 & 0.975 & 0.440\\
& MEDRV & 0.054 & 0.056 & 0.064 & 0.068 & 1.000 & 1.000 & 0.996 & 0.649\\
\multicolumn{1}{l}{DGP1} & PZ2 & 0.059 & 0.057 & 0.059 & 0.072 & 1.000 &
1.000 & 1.000 & 0.950\\
\multicolumn{1}{l}{Gaussian noise} & PZ4 & 0.054 & 0.056 & 0.064 & 0.068 &
1.000 & 1.000 & 1.000 & 0.950\\
& ASJ & 0.019 & 0.013 & 0.002 & 0.000 & 0.529 & 0.501 & 0.517 & 0.020\\
& ABD & 0.062 & 0.082 & 0.074 & 0.052 & 1.000 & 1.000 & 1.000 & 0.967\\
& LM & 0.025 & 0.006 & 0.007 & 0.001 & 0.990 & 0.875 & 0.683 & 0.118\\
& JO & 0.006 & 0.039 & 0.042 & 0.072 & 1.000 & 1.000 & 1.000 & 0.809\\
& LM12 & 0.043 & 0.155 & 0.216 & 0.511 & 1.000 & 0.965 & 0.921 & 0.671\\
& AJL & 0.053 & 0.046 & 0.014 & 0.082 & 0.999 & 0.818 & 0.507 & 0.105\\\hline
&  &  &  &  &  &  &  &  & \\
& BNS & 0.117 & 0.050 & 0.046 & 0.065 & 0.999 & 0.999 & 0.991 & 0.548\\
& CPR & 0.410 & 0.192 & 0.210 & 0.488 & 0.999 & 1.000 & 0.997 & 0.846\\
& MINRV & 0.000 & 0.022 & 0.031 & 0.044 & 0.353 & 0.910 & 0.926 & 0.434\\
& MEDRV & 0.000 & 0.036 & 0.044 & 0.054 & 0.610 & 0.952 & 0.964 & 0.639\\
\multicolumn{1}{l}{DGP1} & PZ2 & 0.997 & 0.374 & 0.172 & 0.088 & 1.000 &
1.000 & 1.000 & 0.935\\
\multicolumn{1}{l}{Student-t noise} & PZ4 & 0.997 & 0.370 & 0.183 & 0.079 &
1.000 & 1.000 & 1.000 & 0.935\\
& ASJ & 0.000 & 0.006 & 0.005 & 0.000 & 0.157 & 0.394 & 0.460 & 0.016\\
& ABD & 1.000 & 0.463 & 0.198 & 0.070 & 1.000 & 1.000 & 1.000 & 0.948\\
& LM & 0.722 & 0.063 & 0.018 & 0.002 & 0.996 & 0.893 & 0.680 & 0.114\\
& JO & 0.208 & 0.074 & 0.056 & 0.084 & 0.997 & 0.996 & 0.997 & 0.798\\
& LM12 & 0.051 & 0.148 & 0.221 & 0.512 & 1.000 & 0.955 & 0.925 & 0.722\\
& AJL & 0.063 & 0.054 & 0.018 & 0.101 & 0.998 & 0.798 & 0.488 & 0.110\\\hline
&  &  &  &  &  &  &  &  & \\
& BNS & 0.000 & 0.050 & 0.063 & 0.058 & 1.000 & 1.000 & 0.994 & 0.560\\
& CPR & 0.000 & 0.133 & 0.189 & 0.468 & 1.000 & 1.000 & 0.998 & 0.827\\
& MINRV & 0.000 & 0.047 & 0.044 & 0.041 & 1.000 & 1.000 & 0.985 & 0.459\\
& MEDRV & 0.000 & 0.056 & 0.048 & 0.051 & 1.000 & 1.000 & 1.000 & 0.619\\
\multicolumn{1}{l}{DGP1} & PZ2 & 0.062 & 0.062 & 0.068 & 0.082 & 1.000 &
1.000 & 1.000 & 0.939\\
\multicolumn{1}{l}{Gaussian} & PZ4 & 0.053 & 0.058 & 0.063 & 0.077 & 1.000 &
1.000 & 1.000 & 0.937\\
\multicolumn{1}{l}{Autocorrelated noise} & ASJ & 0.023 & 0.015 & 0.004 &
0.000 & 0.517 & 0.486 & 0.475 & 0.014\\
& ABD & 0.039 & 0.047 & 0.056 & 0.060 & 1.000 & 1.000 & 1.000 & 0.956\\
& LM & 0.020 & 0.009 & 0.007 & 0.000 & 0.999 & 0.894 & 0.710 & 0.115\\
& JO & 0.013 & 0.040 & 0.048 & 0.086 & 1.000 & 1.000 & 1.000 & 0.797\\
& LM12 & 0.030 & 0.164 & 0.216 & 0.536 & 1.000 & 0.958 & 0.925 & 0.736\\
& ASJL & 0.070 & 0.058 & 0.015 & 0.095 & 0.997 & 0.822 & 0.488 &
0.118\\\hline\hline
\end{tabular}
\end{table}
\end{document}